\documentclass[grl]{AGUTeX}
\usepackage{graphicx}
\usepackage{amsmath}
\pdfoutput=1

\authorrunninghead{TANG ET AL.}
\titlerunninghead{OBSERVATIONS OF WAVES IN DIFFUSION REGIONS}

\authoraddr{X. Tang,
116 Church Street S.E., Minneapolis, MN, 55455, USA.
(xtangphysics@gmail.com)}

\newcommand{\figsquash}{\ifjdraft\baselineskip=5pt\fi}

\begin{document}

\title{THEMIS Observations of the Magnetopause Electron Dif\mbox{}fusion Region: Large Amplitude Waves and Heated Electrons}

\authors{Xiangwei Tang,\altaffilmark{1}
Cynthia Cattell,\altaffilmark{1} John Dombeck,\altaffilmark{1}
Lei Dai,\altaffilmark{1} Lynn B. Wilson III,\altaffilmark{2}
Aaron Breneman,\altaffilmark{1} Adam Hupach \altaffilmark{1} }

\altaffiltext{1}{School of Physics and Astronomy,
University of Minnesota, Minneapolis, Minnesota, USA.}

\altaffiltext{2}{Goddard Space Flight Center, 
Heliospheric Physics Laboratory, Greenbelt, Maryland, USA.}

\begin{abstract}
We present the first observations of large amplitude waves in a well-def\mbox{}ined electron dif\mbox{}fusion region at the sub-solar magnetopause using data from one THEMIS satellite. These waves identif\mbox{}ied as whistler mode waves, electrostatic solitary waves, lower hybrid waves and electrostatic electron cyclotron waves, are observed in the same 12-sec waveform capture and in association with signatures of active magnetic reconnection. The large amplitude waves in the electron dif\mbox{}fusion region are coincident with abrupt increases in electron parallel temperature suggesting strong wave heating. The whistler mode waves which are at the electron scale and enable us to probe electron dynamics in the dif\mbox{}fusion region were analyzed in detail. The energetic electrons ($\sim$30 keV) within the electron dif\mbox{}fusion region have anisotropic distributions with $T_{e\perp}/T_{e\parallel}>1$ that may provide the free energy for the whistler mode waves. The energetic anisotropic electrons may be produced during the reconnection process. The whistler mode waves propagate away from the center of the `X-line' along magnetic f\mbox{}ield lines, suggesting that the electron dif\mbox{}fusion region is a possible source region of the whistler mode waves.\\
\end{abstract}

\begin{article}

\section{Introduction}
Magnetic reconnection is considered to be an important energy conversion process \citep{Dungey1961} that occurs in a variety of plasma environments. At the Earth's magnetopause, it facilitates the entry of solar wind plasma and electromagnetic energy into the magnetosphere. Reconnection sites are regions of strong wave activity covering a broad range of frequencies. Wave modes frequently observed near reconnection sites include the whistler mode (WH) waves \citep{Deng2001,Petkaki2006}, electrostatic solitary waves (ESWs) \citep{Farrell2002,Matsumoto2003}, lower hybrid (LH) waves \citep{Cattell1986,Bale2002}, kinetic $Alfv\acute{e}n$ waves \citep{Chaston2005} and Langmuir/upper hybrid waves \citep{Farrell2002}. The effect of dif\mbox{}ferent wave modes on the reconnection process has been a problem of longstanding interest - for their role in anomalous resistivity, particle acceleration, energy transport and formation of reconnection sites \citep{Huba1977,Labelle1988,Treumann1991,Drake2003}.\\

Observations of the electron dif\mbox{}fusion region (EDR) have been made by Polar at the subsolar magnetopause \citep{Mozer2002PRL}, by Wind in the magnetotail \citep{Oieroset2002} and by Cluster in the magnetosheath \citep{Phan2007}. Recent simulations and observations of EDRs during collisionless antiparallel reconnection in Earth's magnetotail \citep{Ng2011} report that the dif\mbox{}fusion region is characterized by a narrow extended layer containing electron jets. It is shown that the jets in the layer are driven by electron pressure anisotropy $P_{e\parallel}>>P_{e\perp}$ and the anisotropy is responsible for the structure of the EDR \citep{Ng2011}. \cite{Mozer2005JGR} has identif\mbox{}ied EDRs on the basis of the non-zero parallel electric f\mbox{}ield, a large perpendicular electric f\mbox{}ield compared to the reconnection electric f\mbox{}ield, a large electromagnetic energy conversion rate and accelerated electrons, and a topological boundary that separates regions having different $\textbf{E}\times\textbf{B}/B^2$ f\mbox{}lows with thickness of the order of the electron skin depth. In this paper, we concentrate on a specific way to identify the EDR described by \cite{Scudder2012} who report spatially resolved diagnostic signatures of a demagnetized EDR observed by Polar at the Earth's magnetopause. The five dimensionless scalar diagnostics that were used to find the EDR are peak electron thermal Mach numbers $M_{e\perp}\equiv\frac{\mid\mathbf{U}_{e}\mid}{<w_{e\perp}>}>1.5$ where $\mathbf{U}_{e}$ represents electron bulk velocity and $<w_{e\perp}>$ is the electron thermal speed derived from the average perpendicular temperatures, electron temperature anisotropy $An_{e}\equiv\frac{T_{e\parallel}}{<T_{e\perp}>}>7$, calibrated agyrotropy of electron pressure tensor $A\phi_{e}=2\frac{\mid1-\alpha\mid}{(1+\alpha)}>1$ where $\alpha\equiv P_{e\perp,1}/P_{e\perp,2}$, expansion parameters of guiding center theory indicative of demagnetization and strong ($150eV$) increases in electron temperature \citep{Scudder2012}.\\

This paper focuses on the WH waves which are an important candidate for the anomalous resistivity, particle acceleration and heating. WH waves may be driven unstable by superthermal electrons with temperature anisotropies of $T_{e\perp}/T_{e\parallel}>1$ in the magnetosphere \citep{Kennel1966} and current-driven plasma instabilities \citep{Gurnett1976} or energetic electron beams \citep{Zhang1999} in the magnetotail. WH waves are one of the most ubiquitous wave modes observed in space plasmas. Observations of WH waves at the Earth's magnetopause have been made by \cite{Deng2001}. Electron anisotropy, due to compression of the magnetopause or LH drift waves, may be the generation mechanism of WH waves in the magnetopause current sheet \citep{Karimabadi2004}. WH waves in the EDR may play a significant role in the microphysics of reconnection as they are excited on electron scales. It is believed that WH waves in the magnetopause current sheet may affect the instability of the current sheet to reconnection via tearing. The generation of the out-of-plane component of the magnetic f\mbox{}ield is suggested to be a signature of whistler mediated reconnection \citep{Mandt1994}. It has also been suggested that the strongest whistler emissions are observed on the most recently opened magnetospheric flux tubes due to magnetic reconnection \citep{Vaivads2007}. One recent simulation study concludes that WH waves do not control the dissipation processes of reconnection but are generated as a result of the reconnection processes \citep{Fujimoto2008}. In this paper, we present an example of a reconnection event at the sub-solar magnetopause observed by THEMIS. In section 2, we describe the data sets and analysis techniques. In section 3, we show the observations. Finally, we discuss the conclusions of our study in section 4.\\

\section{Data Sets and Analysis}

The THEMIS mission consists of five identically-instrumented spacecraft \citep{Angelopoulos2008}. The Electric F\mbox{}ield Instrument (EFI) measures three components of the electric f\mbox{}ield \citep{Bonnell2008}. The instrument provides continuous coverage at 128 samples/s in survey mode and waveform captures at 8192 samples/s in wave burst mode. The Magnetic Fields Experiment (MFE) includes a Flux Gate Magnetometer (FGM) which measures DC magnetic f\mbox{}ield with a sampling rate of 128 samples/s in the high rate mode or 4 samples/s in the low rate mode \citep{Auster2008} and a Search Coil Magnetometer (SCM) which measures magnetic fluctuations sampled at 8192 samples/s in the burst mode \citep{Roux2008,Contel2008}. Particle data are measured by the Electrostatic Analyzer (ESA) \citep{McFadden2008} and by the Solid State Telescope (SST) \citep{Angelopoulos2008}. The ESA measures plasma over the energy range of a few eV up to 30 keV for electrons and 25 keV for ions. The SST measures the distribution functions of superthermal particles within the energy range from 25 keV to 6 MeV for electrons and 900 keV for ions. Because the measured quasi-static electric f\mbox{}ield component along the spin axis has large uncertainty due to the short boom along the spin axis we use $\mathbf{E}\cdot\mathbf{B}=0$ to determine the electric f\mbox{}ield used to calculate the $\mathbf{E}\times\mathbf{B}/B^2$ velocity. The coordinate systems used in this paper include geocentric solar magnetospheric (GSM) coordinates and f\mbox{}ield-aligned coordinates (FAC). The FAC is def\mbox{}ined in the following way: The positive Z axis points in the direction of the magnetic f\mbox{}ield at the spacecraft's location. The positive X axis lies in the plane of the magnetic f\mbox{}ield line passing through the spacecraft's location, perpendicular to the Z axis, and points inwards (towards the inside of the f\mbox{}ield line). The positive Y axis completes the orthogonal right-handed system. Waveforms are analyzed dynamically in time and frequency using Morlet wavelet transform \citep{Torrence1998}. The wave vector is determined using Minimum Variance Analysis (MVA) \citep{Khrabrov1998} on bandpass filtered three magnetic f\mbox{}ield components.\\

\section{Observations}

\begin{figure}[ht]
\centering
\noindent
\includegraphics[width=30pc]{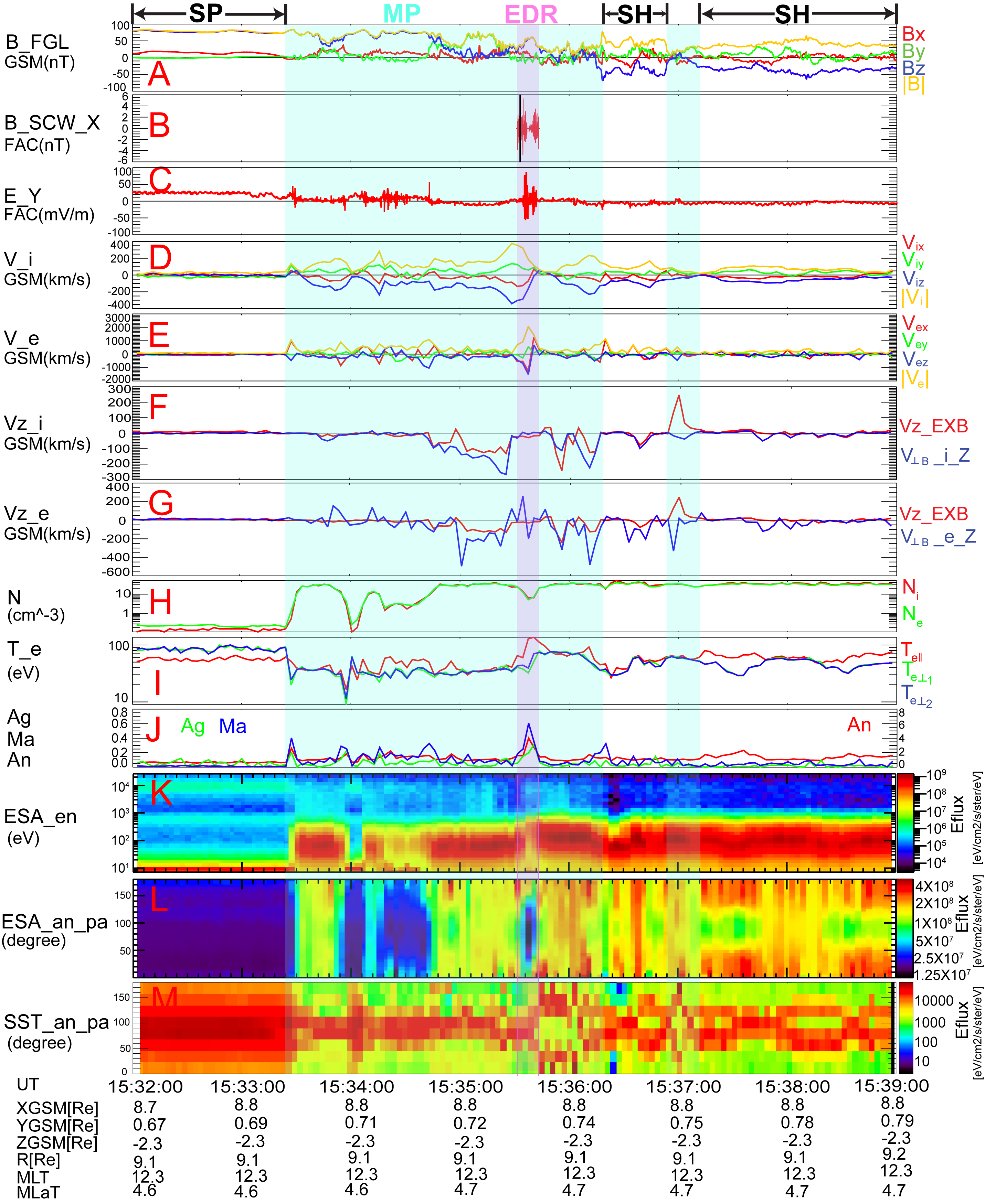}
\caption{\figsquash{A reconnection event at the sub-solar magnetopause observed by THEMIS-E on August 27, 2009. (A): 4 samples/s magnetic f\mbox{}ield data in GSM. (B): perpendicular X component of the burst magnetic f\mbox{}ield at 8192 samples/s in FAC. (C): perpendicular Y component of the electric f\mbox{}ield at 128 samples/s in FAC. (D) and (E): ion and electron bulk f\mbox{}low velocity in GSM, respectively. (F) and (G): comparisons of GSM-Z component of the $\mathbf{E}\times\mathbf{B}/B^2$ velocity with the GSM-Z component of ion (F) and electron (G) perpendicular f\mbox{}low velocity with respect to the ambient magnetic f\mbox{}ield, respectively. (H): ion and electron densities. (I): electron temperatures. (J): electron agyrotropy and Mach number (scale to the left) and temperature anisotropy (scale to the right). (K): differential energy flux for electrons measured by ESA. (L) and (M): electron pitch angle spectra for lower engery electrons measured by ESA and higher energy electrons measured by SST, respectively.}}
\label{Fig1}
\end{figure}

Figure \ref{Fig1} shows a 7-min interval of the f\mbox{}ield and plasma observations made by probe E of the THEMIS mission on August 27, 2009. The boundary normal direction (determined from MVA on the ambient magnetic f\mbox{}ield) was [0.99, 0.015, -0.12] in GSM coordinates and almost identical to the GSM-X direction, consistent with the spacecraft being near the sub-solar point (indicated by the position parameters at the bottom of Figure \ref{Fig1}). The spacecraft travels from the outer magnetosphere (SP) through the magnetopause (MP, indicated by two light green shaded bands) into the magnetosheath (SH). The purple shaded band shows an $\sim$12-sec interval of magnetic burst data capture.\\

The magnetopause crossing can be seen in the change in Bz from positive to negative in Panel A of Figure \ref{Fig1}. The differential energy flux of electrons in Panel K shows that in the magnetosphere, where Bz was positive, high-energy electrons were encountered; while in the magnetosheath, where Bz was negative, lower-energy electrons were measured. The spacecraft passed from the lower plasma density mangetosphere to the higher density (factor of 100) magnetosheath via a region of mixed magnetosheath/magnetospheric plasmas comprising the low latitude boundary layer as shown in Panel H. This observation is evidence for the transport of solar wind plasma across the magnetopause. The presence of accelerated plasma f\mbox{}low is seen through the magnetopause current sheet as shown in Panels D and E. The spacecraft crossed the magnetopause south of the separator, as suggested by the negative GSM-Z component of ion f\mbox{}low velocity and the result of Walen test \citep{Sonnerup1981}. Based on the prediction of the magnetopause reconnection model being a rotational discontinuity \citep{Paschmann1979}, the Walen test \citep{Sonnerup1981} states that the observed f\mbox{}low velocity change between a point in the magnetopause and a reference point in the adjacent magnetosheath equals the predicted modified Alfven velocity change. The angle deviations between the observations and the prediction are almost $180^\circ$ for this event. This anti-parallel relation indicates that the spacecraft crossed south of the separator \citep{Sonnerup1981}. The encounter with the magnetopause current sheet is associated with fast ion jetting consistent with the Walen relation and fast electron f\mbox{}lows, indicating that reconnection is occurring. Magnetic reconnection is generally considered to be the primary mechanism through which transport of plasma and energy across the magnetopause occurs.\\

\subsection{Identification of EDR}

Enhanced wave activity can be seen associated with the magnetopause crossing from Panels A (ambient magnetic f\mbox{}ield), B (burst magnetic f\mbox{}ield) and C (electric fluctuations). We note that the electric fluctuations maximize during the magnetic burst interval. During this interval, electron f\mbox{}low speed (Panel E), anisotropy, agyrotropy and Mach number (Panel J) also maximize. These enhanced amplitudes are coincident with abrupt increases in electron parallel temperature $T_{e\parallel}$ shown in Panel I suggesting strong wave heating. As will be discussed in more detail in next section, the observed intense waves may provide the observed electron heating. All these features, along with the fact that the electron perpendicular f\mbox{}low velocity is not consistent with the $\mathbf{E}\times\mathbf{B}/B^2$ velocity during the magnetic burst interval (Panel G), provide evidence for the detection of an EDR. Another feature is the density depletion shown in Panel H (purple shaded band) which might indicate the center of the EDR. These signatures are consistent with the simulation and observations of \cite{Scudder2012}. Panels L and M show the electron pitch angle spectra for lower engery electrons measured by ESA and higher energy electrons measured by SST, respectively. The magnetic f\mbox{}ield (Panel A) and electric f\mbox{}ield (Panel C) fluctuations enhance in the magnetopause boundary layer and in the magnetosheath with f\mbox{}ield-aligned and counter-streaming lower energy electrons as shown in Panel L. However, the higher energy electron pitch angle (Panel M) enhances around $90^\circ$ in the boundary layer and in the magnetosheath, especially during the purple shaded magnetic burst interval. Distinct from the electron distributions in the boundary layer and in the magnetosheath, the higher energy electron pitch angle enhances at $0^\circ$ and $180^\circ$ and the lower energy electrons are more isotropic near the current sheet center around 15:36:00 UT and 15:37:00 UT.\\

\subsection{Observations of waves}
\begin{figure}
\noindent
\includegraphics[width=21pc]{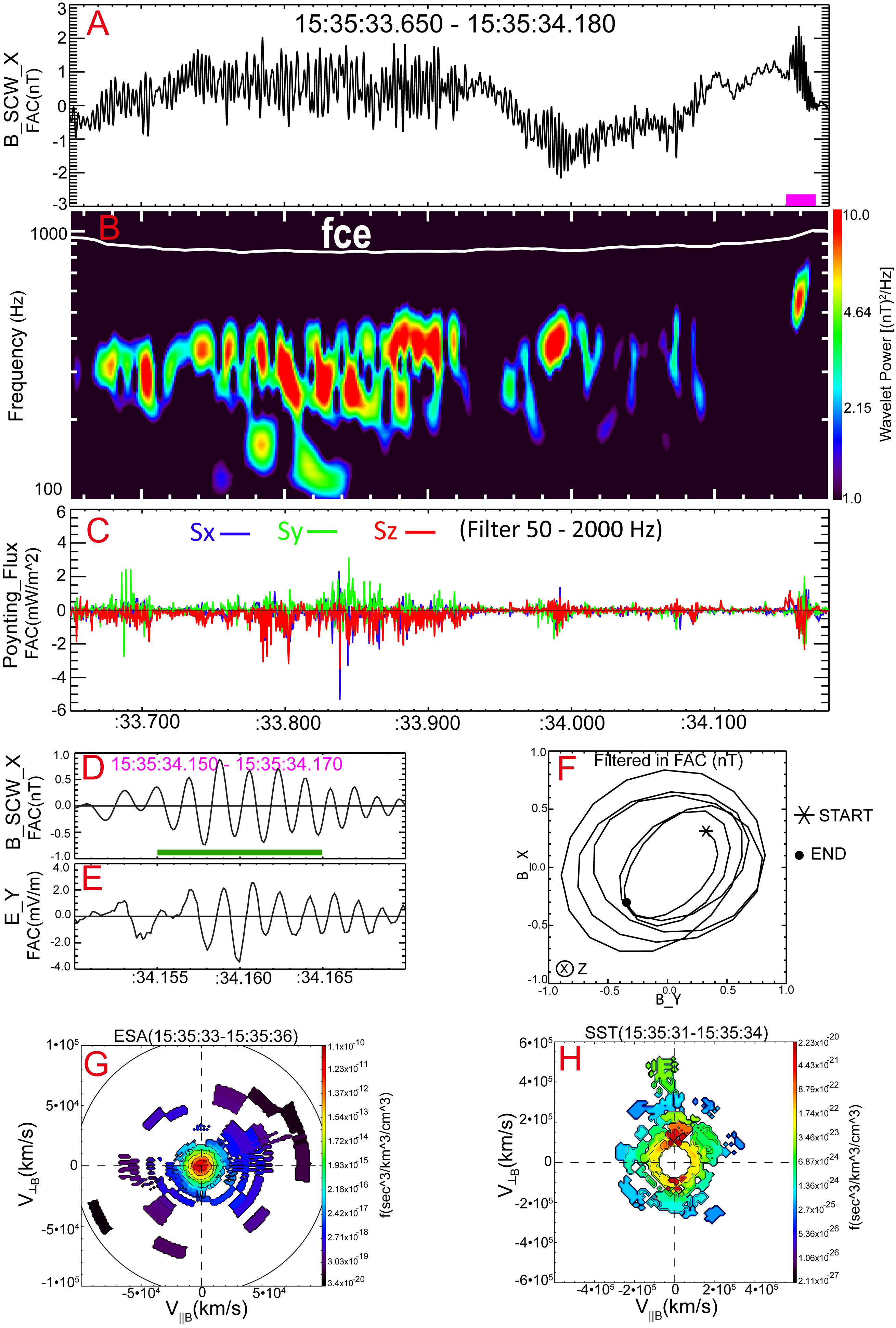}
\caption{\figsquash{Example of WH waves within the EDR. (A), (B) and (C): a 0.53-sec interval in FAC of perpendicular X component of the burst magnetic f\mbox{}ield, associated Wavelet power spectrum and whistler Poynting flux, respectively. (D) and (E): expanded views of the filtered whistler waveforms of the perpendicular X component of the burst magnetic f\mbox{}ield and the perpendicular Y component of the electric f\mbox{}ield over the time interval indicated by the purple bar in (A). (F): hodogram of the filtered burst magnetic f\mbox{}ield waveforms in FAC for the interval indicated by the green bar in (D). The black star and dot mark the beginning and ending of the wave f\mbox{}ield, respectively. (G) and (H): distribution functions of lower energy electrons (up to 20 keV) measured by ESA and higher energy electrons (20-700 keV) measured by SST observed at times close to the WH waves, respectively. The horizontal axis is parallel to the ambient magnetic f\mbox{}ield and the bulk velocity def\mbox{}ines the plane.}}
\label{Fig2}
\end{figure}

Figure \ref{Fig2} shows an example of the identif\mbox{}ied WH waves at the time indicated by a black vertical line in Panel B of Figure \ref{Fig1}. It can be seen from Panels A and B of Figure \ref{Fig2} that the waves have frequencies from 0.1 to 0.6 $f_{ce}$ (electron cyclotron frequency) with amplitudes up to 3 nT (peak-peak). As can be seen from Panel C, the wave Poynting flux is mostly anti-parallel to the ambient magnetic f\mbox{}ield. The wave vector ($k=[-0.23,-0.09,0.97]$ in FAC), determined from MVA on bandpass filtered wave magnetic f\mbox{}ield, is nearly along the background magnetic f\mbox{}ield. The wave propagation angle with respect to the ambient magnetic f\mbox{}ield $\theta_{kB}$ is determined as $\sim166^\circ$ since the Poynting flux is mostly anti-parallel to the magnetic f\mbox{}ield. An expanded view of the WH waves can be seen in Panels D and E which respectively show the filtered (200-2000 Hz) waveforms of the burst magnetic and electric f\mbox{}ield data over the time interval indicated by the purple bar in Panel A. Panel F shows that the WH waves are circularly right-handed polarized with respect to the ambient magnetic f\mbox{}ield as expected. The electron distribution functions observed at times close to and/or concurrently with the WH waves are shown in Panels G and H. The lower energy electrons ($\sim$100 eV) shown in Panel G have anisotropic distributions with $T_{e\perp}/T_{e\parallel}<1$. However, the energetic electrons ($\sim$30 keV) shown in Panel H have anisotropic distributions with a larger population moving in perpendicular direction. Broad-banded emissions with strong electric ($\sim$10 mV/m) and magnetic ($\sim$40 nT) f\mbox{}ield fluctuations below ion cyclotron frequency ($\sim$1 Hz) and electric ($\sim$30 mV/m) and magnetic ($\sim$20 nT) f\mbox{}ield fluctuations below LH frequency ($\sim$30 Hz) are also detected during this magnetopause crossing (not shown). These intense wave emissions may provide the observed electron heating associated with the magnetopause crossing.\\

\begin{figure}
\noindent
\includegraphics[width=21pc]{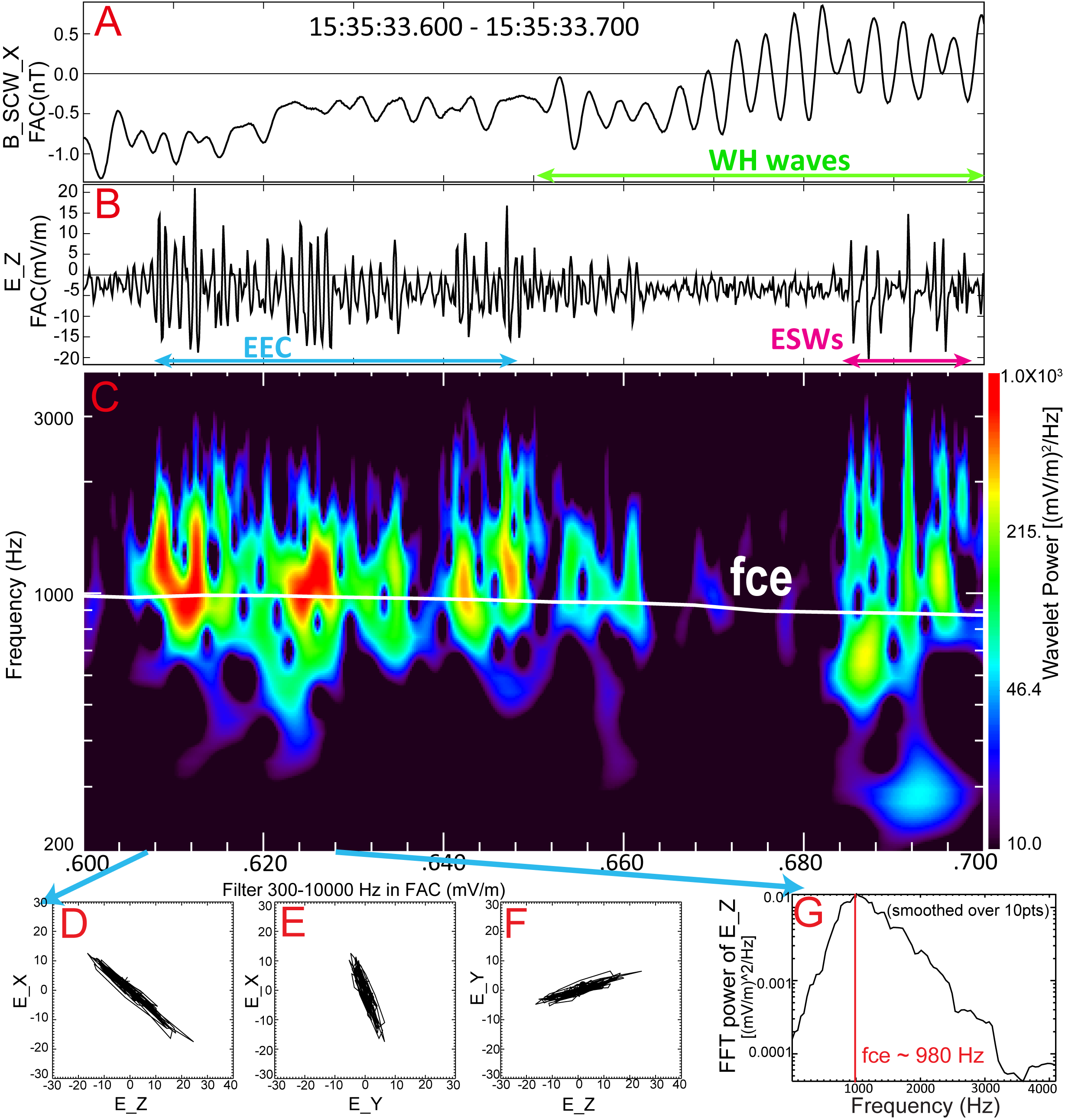}
\caption{\figsquash{Example of electrostatic waves within the EDR. (A), (B) and (C): a 0.1-sec interval of the perpendicular X component of the burst magnetic f\mbox{}ield, the parallel Z component of the electric f\mbox{}ield waveform capture and the associated Wavelet power spectrum of the parallel electric f\mbox{}ield in FAC, respectively. (D), (E) and (F): hodograms of the electric f\mbox{}ield waveforms with an interval indicated by the light blue arrows below (C). The definitions of the black star and dot on hodograms are the same as those in Figure \ref{Fig2}. (G): Fourier wave power vs. frequency with the same time interval as the hodograms.}}
\label{Fig3}
\end{figure}

Figure \ref{Fig3} shows examples of electrostatic waves at a time preceding the waves in Figure \ref{Fig2} by 0.05 seconds. WH waves at a frequency of $\sim0.3f_{ce}$ are observed in the magnetic fluctuations in Panel A from 15:35:33.650 UT to 15:35:33.700 UT. This time interval overlaps with that of the ESWs, suggesting a possible coupling of WH waves and ESWs. ESWs (up to 30 mV/m) indicated by the magenta arrows in Panel B have a broad spectrum which extends from 200 Hz to 3000 Hz shown in Panel C. The high-frequency electrostatic waves (up to 35 mV/m) labeled by light blue arrows in Panel B have power that peaks at $f_{ce}$, which can be seen from both the wavelet power spectrum in Panel C and the Fourier power in Panel G. During this time interval, there is no power in the wave magnetic f\mbox{}ield at $f_{ce}$ (not shown). This wave mode is linearly polarized, as shown in the hodograms in Panels D, E, and F in FAC with an interval indicated by the light blue arrows below Panel C. Occasionally these high-frequency emissions are seen associated with clear harmonics, possibly suggesting electrostatic electron cyclotron (EEC) waves.\\ 

\section{Discussion and Conclusions}
We have presented the first observations of intense waves in the EDR in a sub-solar magnetopause reconnection region. The identification of the EDR in this event is based on the occurrence of signatures of strong electron heating, large electron thermal anisotropy, agyrotropy and Mach number and electron velocity not consistent with the $\mathbf{E}\times\mathbf{B}$ velocity, consistent with the simulation and observations of \cite{Scudder2012}. The lower energy electrons ($\sim$100 eV) with anisotropic distributions of $T_{e\perp}/T_{e\parallel}<1$ within the EDR may have been heated by the observed waves with frequency below the LH frequency, consistent with the suggestion that LH waves lead to electron heating in the parallel direction \citep{Cairns2005}.\\

We identif\mbox{}ied intense WH waves inside the EDR. This is inconsistent with reported simulation results that indicated WH waves are only driven downstream of an EDR \citep{Fujimoto2008}. The WH waves seen by THEMIS in the EDR propagate almost anti-parallel to the ambient magnetic f\mbox{}ield and the Poynting flux indicates that the WH waves propagate away from the center of the `X-line' along magnetic f\mbox{}ield lines. The observed electron temperature anisotropy of $T_{e\perp}/T_{e\parallel}>1$ for energies above 20keV may be the source of free energy for the generation of the WH waves. The energetic electron anisotropy may be produced by adiabatic heating in the perpendicular direction as the locally intensified magnetic f\mbox{}ield can accelerate electrons in the perpendicular direction \citep{Fujimoto2008}. On the f\mbox{}ield lines directly connected to the EDR, the energetic electron anisotropy may also be due to high energy f\mbox{}ield-aligned electrons (accelerated by the reconnection process) being lost to the magnetosheath \citep{Stenberg2005}. WH waves can scatter the electrons in pitch-angle distribution and relax the temperature anisotropy. Studies of large amplitude whistlers in the inner magnetosphere have provided evidence for rapid scattering and/or energization \citep{Cattell2008}. WH waves may play a significant role in the microphysics of reconnection through the enabling of a current sheet instability, the decoupling of electrons, the acceleration and heating of particles, and the transport of energy away from the reconnection region.\\

A possible coupling of EEC waves and ESWs with WH waves is often seen during magnetopause reconnection. The growth of the electrostatic waves may reduce the electron temperature anisotropy and reduce the growth rate of WH waves. The physics of wave coupling process is important to understand the effect of wave-wave interactions on the reconnection process and will be investigated in a future study. This study provides further evidence that the plasma waves can play a significant role in the microphysics of magnetic reconnection at the Earth's magnetopause.\\

\begin{acknowledgments}
At the University of Minnesota, this work was supported by NNX08AF28 and a contract from APL for the development of RBSP/EFW. The authors acknowledge NASA contract NAS5-02099 and V. Angelopoulos for use of data from the THEMIS Mission, specifically: J. W. Bonnell and F. S. Mozer for use of EFI data; D. Larson and R. P. Lin for use of SST data; C. W. Carlson and J. P. McFadden for use of ESA data; A. Roux and O. LeContel for use of SCM data; and K. H. Glassmeier, U. Auster and W. Baumjohann for the use of FGM data provided under the lead of the Technical University of Braunschweig and with financial support through the German Ministry for Economy and Technology and the German Center for Aviation and Space (DLR) under contract 50 OC 0302. The authors are grateful for discussion and comments from S. Thaller, K. Kersten and C. Colpitts.\\
\end{acknowledgments}


\clearpage

\clearpage

\end{article}

\end{document}